\documentclass[preprint,showpacs,preprintnumbers,amsmath,amssymb]{revtex4}

\usepackage{graphicx}% Include figure files
\usepackage{dcolumn}% Align table columns on decimal point
\usepackage{bm}% bold math

\begin{document}

%\preprint{APS/123-QED}

\title{On the theory of pseudogap anisotropy in the cuprate
       superconductors}% Force line breaks with \\

\author{V.M. Loktev}
 \altaffiliation{Electronic address: vloktev@bitp.kiev.ua}
\affiliation{%
Bogolyubov Institute for Theoretical Physics, 
Metrologicheskaya str. 14-b, Kiev-143, 03143 Ukraine}%

\author{V. Turkowski}
\altaffiliation{Electronic address: turk@physics.georgetown.edu}
\affiliation{Department of Physics, Georgetown University,
             Washington, D.C. 20057}%

\date{\today}% It is always \today, today,

\begin{abstract}
We show by means of the theory
of the order parameter phase fluctuations
 that the temperature of the ``closing'' 
(or ``opening'') of the gap (and pseudogap)
in the electron spectra of superconductors
with anisotropic order parameter actually
takes place within a finite temperature range.
Every Fourier-component of the order parameter
has its own critical temperature.
\end{abstract}


\maketitle

It is well known that the it is not easy to build a
self-consistent theory of the copper oxide high-temperature
superconductors (HTSCs) due to the necessity to take into account
different properties of the cuprates, in particular
strong electron correlations, low dimensionality
of the electronic and magnetic properties, anisotropy of the order
parameter, pseudogap presence, disorder etc.
It is extremely difficult to include
all these properties into the theory self-consistently.
The choice of the properties is usually dictated by
the aim of studies. Below we make an attempt
to show that it is possible to explain such an unusual
phenomenon as smooth disappearing (``closing'') of the
pseudogap along the Fermi surface arcs from 
nodal points to M-points ($(0,\pi )$ or $(\pi ,0)$) of
the Brillouin zone in the momentum space.

It has been proved that the description
of the superconductivity
in the 2D metals with arbitrary carrier density
requires not two, as usual, but three self-consistent
equations (see review \cite{Loktev1}).
Two of them are well known. The first one is the gap
equation, which defines the order parameter. It can be
written in the following form in the case of a separable
interaction \cite{Schrieffer}:

\begin{equation}
\Delta_{\bf k}=\frac{V}{N}\gamma_{l} ({\bf k})
\sum_{\bf q}\gamma_{l} ({\bf q})
\frac{1}{2\pi}
\int_{-\infty}^{+\infty}
\frac{d\varepsilon}{\exp(\varepsilon /T)+1}
Im Tr [{\hat \tau}_{1}{\hat G}_{\bf q}(\varepsilon )] ,
\label{gapequation}
\end{equation}
where
\begin{equation}
{\hat G}_{\bf q}(\omega_{n})=
\frac{1}{i\omega_{n}-\xi ({\bf k}){\hat \tau}_{1}
-\Delta_{\bf k}{\hat \tau}_{3}}, 
\label{GF}
\end{equation}
is the Matsubara fermion Green's function in the Nambu representation,
and 
\begin{equation}
\xi ({\bf k})= \frac{W}{2}-\frac{W}{4}(cos(k_{x}a)+cos(k_{y}a))-\mu
\label{xi}
\end{equation}
is a spectrum of one-particle fermi excitations on the square lattice
with the constant $a$ and with the bandwidth $W$. 
The spectrum energy is measured
with respect to the chemical potential $\mu$.
The order parameter in different pairing channels is
$\Delta ({\bf k})=\Delta_{l}(T)\gamma_{l}({\bf k})$,
where $l$ stands for the symmetry of the order parameter.
In the $s$-wave pairing channel we choose
$\gamma_{s}({\bf k})=1$, in the $p$-wave channel 
$\gamma_{p}({\bf k})=sin(k_{x}a)$, and, finally,
in the related to the HTSCs $d$-wave pairing channel
$\gamma_{d}({\bf k})= cos(ak_{x})-cos(ak_{y})$.
$\omega_{n}=(2n+1)\pi T$ are the Matsubara fermi-frequency,
${\hat \tau}_{i}$ are the Pauli matrices, and $V$ is the parameter
which characterizes the fermion attraction
(the separable pairing potential is chosen
in the following form: 
$V_{l}({\bf k},{\bf q})=V\gamma_{l}({\bf k})\gamma_{l}({\bf q})$).

The second equation can be written as:
\begin{equation}
n_{f}=1-\frac{1}{N}
\sum_{\bf k}
\frac{1}{\pi}
\int_{-\infty}^{+\infty}
\frac{d\varepsilon }{\exp(\varepsilon /T)+1}
Im Tr [{\hat \tau}_{3}{\hat G}_{\bf k}(\varepsilon )].
\label{numberequation}
\end{equation}
This equation allows to connect the number of
the mobile (doped) carriers in the conducting band 
with the dispersion law (\ref{xi}) and the chemical potential
(usually, holes are the carriers in the copper oxide HTSCs).
Generally, the value of the chemical potential is not equal
to the value of the Fermi energy (see \cite{Loktev1}).

After the integration over the energy $\varepsilon$
and by using (\ref{GF}) and (\ref{xi}),
equations (\ref{gapequation}) and (\ref{numberequation})
can be transformed to a more familiar form:
\begin{equation}
1=V
\int \frac{d^{2}k}{(2\pi )^{2}}
\gamma_{l}({\bf k})^{2}
\frac{\tanh [E({\bf k})/(2T)]}{2E({\bf k})};
\label{gapequation0}
\end{equation}
\begin{equation}
n_{f}=\int \frac{d^{2}k}{(2\pi )^{2}}
\left[ 1-
\frac{\xi ({\bf k})}{E({\bf k})}
\tanh\left(\frac{E({\bf k})}{2T}\right)
\right] ,
\label{numberequation0}
\end{equation}
where $E({\bf k})=\sqrt{\xi^{2}({\bf k})+|\Delta ({\bf k})|^{2}}$
is the excitation energy of the quasi-particles in the
superconducting state. The function  $E({\bf k})$
is equal to zero at so called nodal points (with 
$|k_{x}|=|k_{y}|=|k_{F}|$, where ${\bf k}_{F}$ is the Fermi momentum).

The system (\ref{gapequation0}) and (\ref{numberequation0})
is the self-consistent system of equation of the BCS theory.
Its solution in the 2D case describes the temperature dependence of the
gap (amplitude of the order parameter) and
of the chemical potential at given value of $n_{f}$.
It is impossible to estimate the correct value
of the critical temperature from this system, since
the critical temperature in the 2D case is equal to zero
due to the long-wave fluctuations of the phase
of the order parameter \cite{Patashinskii}.
However, there is another phase transition in the 2D systems -
the Berezinskii-Kosterlitz-Thouless (BKT) phase transition
with the critical temperature $T=T_{BKT}$ at which the correlation
of the order parameter changes its space dependence
from the exponential (at $T>T_{BKT}$) to the power law (at $T<T_{BKT}$).
This 
transition was studied most completely in the case of the spin $XY$-model
with the following Hamiltonian \cite{Izyumov}:
$H_{XY}=\frac{J}{2}\sum_{n,{\tilde n}}(\theta_{n}-\theta_{\tilde n})^{2}$,
where $J$ is the exchange coupling and $\theta_{n}$, $\theta_{\tilde n}$
are the phases of the spin vectors on the nearest sites
${\bf n}$ and ${\bf {\tilde n}}$. 
The critical temperature of the BKT transition is defined by the following 
equation:
\begin{equation}
T_{BKT}=\frac{\pi}{2}J
\label{TBKTXY}
\end{equation}

The order parameter of the superconducting metal is a complex
(two-component) function, and usually it can be approximated
as following (see, for example \cite{Loktev2}):
$\Phi ({\bf r}_{1},{\bf r}_{2})\simeq \Delta ({\bf r})
e^{i\theta ({\bf R})}$,
where ${\bf r}={\bf r}_{1}-{\bf r}_{2}$ and 
${\bf R}=({\bf r}_{1}+{\bf r}_{2})$ are the relative coordinate
and the center of mass coordinate of the pair, correspondingly.
In the case of the long-wave approximation the kinetic part of 
the thermodynamic potential has the form of $H_{XY}$,
but in this case the superconducting rigidity
plays the role of the exchange parameter.
This parameter is a function of $T_{c}\equiv T_{BKT}$, 
$\mu$ and $\Delta_{l}(T_{c})$.
Equation
\begin{equation}
T_{c}=\frac{\pi}{2}J(T_{c},\mu ,\Delta_{l} (T_{c}))
\label{TBKT}
\end{equation}
makes the system of equations 
(\ref{gapequation0}), (\ref{numberequation0}) and (\ref{TBKT})
closed. This system of equations allows us to find the
gap, the chemical potential and the critical temperature
$T_{c}$.

The expression for the function $J(T_{c},\mu ,\Delta_{l} (T_{c})$
for the case with an anisotropic order parameter can be found
in the complete analogy with the isotropic $s$-case \cite{Gusynin1}
(see also \cite{Loktev1}). It has the following form:
$$
J(T_{c},\mu ,\Delta_{l} (T_{c}))
=\frac{W}{16}T_{c}
\sum_{n=-\infty}^{\infty}
\frac{d^{2}k}{(2\pi)^{2}}Tr\left[ {\hat \tau}_{3}G_{\bf k}(i\omega_{n})
e^{i\delta\omega_{n}{\hat \tau}_{3}}
+\frac{W}{8}{\bf k}^{2}
G_{\bf k}^{2}(i\omega_{n})\right]
$$
$$
=\frac{W}{16}\left[n_{f}
-\frac{W}{16T_{c}}
\int \frac{d^{2}k}{(2\pi)^{2}}{\bf k}^{2}
\frac{1}{\cosh^{2}(E({\bf k})/(2T_{c}))}\right]; (\delta \rightarrow 0).
$$
This expression together with (\ref{TBKT}) defines the equation for the
critical temperature of the 2D superconducting metal with
arbitrary carrier density:
\begin{equation}
T_{c}=\frac{\pi}{32}W\left[ n_{f}-\frac{W}{16T_{c}}
\int \frac{d^{2}k}{(2\pi)^{2}}{\bf k}^{2}
\frac{1}{\cosh^{2}(E({\bf k})/(2T_{c}))}\right] .
\label{TBKT0}
\end{equation}

The solution of the system of self-consistent
equations (\ref{gapequation0}), (\ref{numberequation0}), and (\ref{TBKT0})
can be found analytically in the $s$-wave case with quadratic dispersion
of the quasi-particles,
$\xi ({\bf k})\sim {\bf k}^{2}$.
In this case $\Delta_{l}(0) \sim \sqrt{n_{f}}, T_{c}\sim n_{f}$
and $\mu$ is always negative when the fermion density
$n_{f}\rightarrow 0$. In the case case
of more general dispersion (see, for example \cite{Patashinskii}),
and, moreover, in the case of the anisotropic order parameter,
the solution can be found only numerically.
This solution gives us the dependencies of $T{c}$, $\mu$
and $\Delta_{l}(T_{c})$ on $n_{f}$.
On the other hand, it is possible to find the
amplitude $\Delta_{l}(T)$ as a function of $T$
from (\ref{numberequation}) and (\ref{gapequation0}).
The solution of these equations at $\Delta_{l}(T)=0$ 
gives us the carrier density
dependence of the critical mean-field temperature $T_{c}^{MF}$
in the $l$-pairing channel. This temperature does not correspond
to any observable phase transition, since
there are no phase transitions in the 2D systems, except
$BKT$-transition as it was mentioned above \cite{Foot}.
This transition is a superconducting phase transition
in the metal with inter-fermion attraction,
despite there is no general spontaneous symmetry breaking in all the system
\cite{Gusynin1} (for details, see \cite{Loktev1}).

The solution of the system of equations (\ref{gapequation0}), 
(\ref{numberequation0}), and (\ref{TBKT0})
in the $s$-wave pairing channel is presented in Figs.1 and 2.
It is important to note that the system of equations
can be analyzed analytically in this case. It is easy to see the
complete symmetry of the solutions with respect to the point
$n_{f}=1$ (the case of half-filling), 
what was already mentioned in \cite{Gorbar,Gusynin2}.
The two-particle (local) bound states exist at any value
of $V/W$ at small values of $n_{f}$ (or $2-n_f$, when we consider
the hole pairing). In other words, there is no threshold value
of the coupling for the bound state formation in the $s$-wave
pairing channel in the 2D case. In this case the chemical
potential is negative at any value of $V/W$ and small enough $n_{f}$,
what indicates the crossover from Bose-Einstein condensation regime
to BCS superconductivity  with carrier density increasing
(or coupling decreasing)\cite{Randeria1,Randeria2}.
As it follows also from these Figures, the following
inequality is always correct $T_{c}<<T_{c}^{MF}$, and the
carrier density dependencies are close to those found
analytically in the case of the quadratic dispersion:
$T_{c}^{MF} \sim \sqrt{n_{f}}, T_{c}\sim n_{f}$ in the low 
carrier density limit.
The canonical BCS relation $2\Delta_{s}(0)/T_{c}^{MF}=3.52$
holds at any carrier density except very low values of $n_{f}$
(see Fig.3), but the relation $2\Delta_{s}(0)/T_{c}$ is increasing
with carrier density decreasing, since it is easy to see that
$\Delta_{s}(0)/T_{c}\sim \sqrt{n_{f}}$ at small $n_{f}$.

In the cases of the anisotropic $p$-wave and $d$-wave pairing
(Figs.4-6 and 7-9, correspondingly)
the behavior of the superconducting parameters with doping
is qualitatively the same. The only difference is that the crossover
to superfluidity takes place only above some critical values of $V$.
The local pairs with non-zero orbital momentum can be formed only
when the attraction is strong enough \cite{Gaididei,Kagan}.
In the cases of the moderate or low attraction the chemical
potential of the system is positive at any value of $n_{f}$, 
therefore only Cooper $p$- and $d$-wave pairs can exist in this case.
The chemical potential in this case practically 
coincides with the Fermi energy. As it follows from Figures
5 and 8, the relation $T_{c}<<T_{c}^{MF}$ holds also in these cases,
and the ratios $2\Delta_{p}(0)/T_{c}^{MF}$ and $2\Delta_{d}(0)/T_{c}^{MF}$
are even higher in comparison with the $s$-wave case, what
shows that the isotropic condensate
is more stable with respect to the thermal fluctuations.
The region between $T_{c}$ and $T_{c}^{MF}$, where
the superconducting fluctuations are incoherent
and rapidly decay and the order parameter modulus is finite,
should be interpreted as the pseudogap region.
Above $T_{c}^{MF}$ (which in some papers (for example, in \cite{Tallon,Naqib}) 
is called the temperature of the decay of superconducting fluctuations 
$T_{scf}$) the gap in the superconducting spectrum disappears,
while there is no any phase transition at the point $T_{c}^{MF}$
or around it.

There is another principal difference between $T_{c}$ and $T_{c}^{MF}$
due to the anisotropy of the electron spectra in the superconducting phase.
The temperature $T_{c}$ is critical temperature of the phase transition, 
and therefore it is unique for all the system and for all its excitations.
This temperature can be measured as the temperature below which
the resistivity goes to zero, or by the Meissner effect, for example.
On the other hand, this statement can not be used with respect
to $T_{c}^{MF}$. This temperature is not the critical temperature
of any phase transition in the 2D system, similarly to the case
with the $s$-wave pairing. The important question is whether
the critical temperature of the formation of different
Fourier-components or the order parameter is unique.
Even the temperature $T_{c}$ can have such values
that the following rations are true 
at different directions of the momentum ${\bf k}$:
$T_{c}\leq \Delta ({\bf k})$, or $T_{c}\ge \Delta ({\bf k})$,
including $T_{c}\gg \Delta ({\bf k})$.
More important question is whether the order parameter modulus
forms at any value of ${\bf k}$.  
A simple (and, in some sense qualitative) answer 
to this question can be given by using the canonical BCS ratio
rewritten in the following form:
\begin{equation}
\frac{2\Delta_{l}(T=0)}{T_{c}^{MF}}\rightarrow 
\frac{2\Delta_{l}(T=0)|\gamma_{l}({\bf k})|}{T_{c}^{MF}(\bf k)},
\label{2Delta_T}
\end{equation}
which shows that every Fourier-component of the anisotropic
order parameter has its own ``closing'' critical temperature.
It should be emphasized that this temperature does not
correspond to any phase transition.

The relation (\ref{2Delta_T}) is an estimation
which actually defines the temperature
$T_{c}^{MF}({\bf k})=T_{c}^{MF}|\gamma_{l}({\bf k})|$ 
(or $T_{scf}({\bf k})$). 
As it follows from this relation,
the gaped quasi-particle spectra of the 2D superconductor
with anisotropic order parameter shows its momentum dependence also when 
the temperature changes. It is important that the gap disappears
continuously from the nodal point, where $\gamma_{l}({\bf k})=0$
at any $T$, to the $M$-points,
where the gap and the corresponding temperature $T_{c}^{MF}$
are maximal, with temperature increasing. 
The gapless (or, actually ``pseudogapless'')
character of the spectra will cover larger parts of the Fermi surface
with temperature increasing at any values of ${\bf k}$,
at which $T=T_{c}^{MF}({\bf k})$.
This behavior was observed in the ARPES experiments (see recent review
\cite{Damascelli,Norman} and references therein). 

Thus, it is shown that the self-consistent study of the
phase fluctuations of the superconducting order parameter
allows one to describe qualitatively the anisotropy of the pseudogap
and its disappearance within a rather wide range of temperatures.

\section*{Acknowledgments}

One of us (V.M.L.) would like to thank
H.~Beck, Yu.B.~Gaididei and V.P.~Gusynin 
for stimulating discussions of the results.
The work was partially supported by 
The Swiss National Science Foundation
(SCOPES-project 7UKPJ062150.00/1).

\newpage

\begin{figure}[h!]% fig. 1
\centering
\includegraphics[width=2.0in,height=3.0in,angle=270]{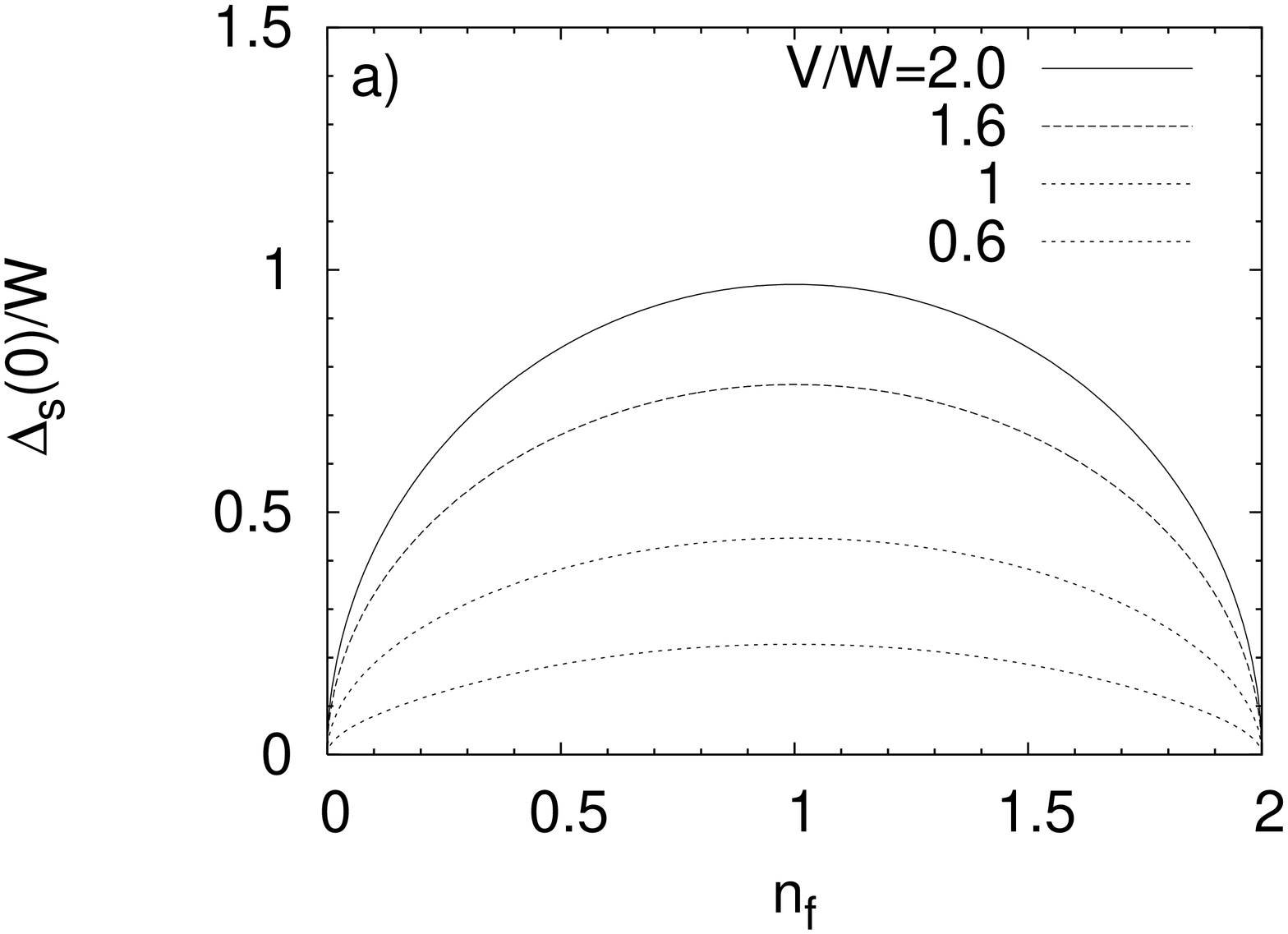}
\includegraphics[width=2.0in,height=3.0in,angle=270]{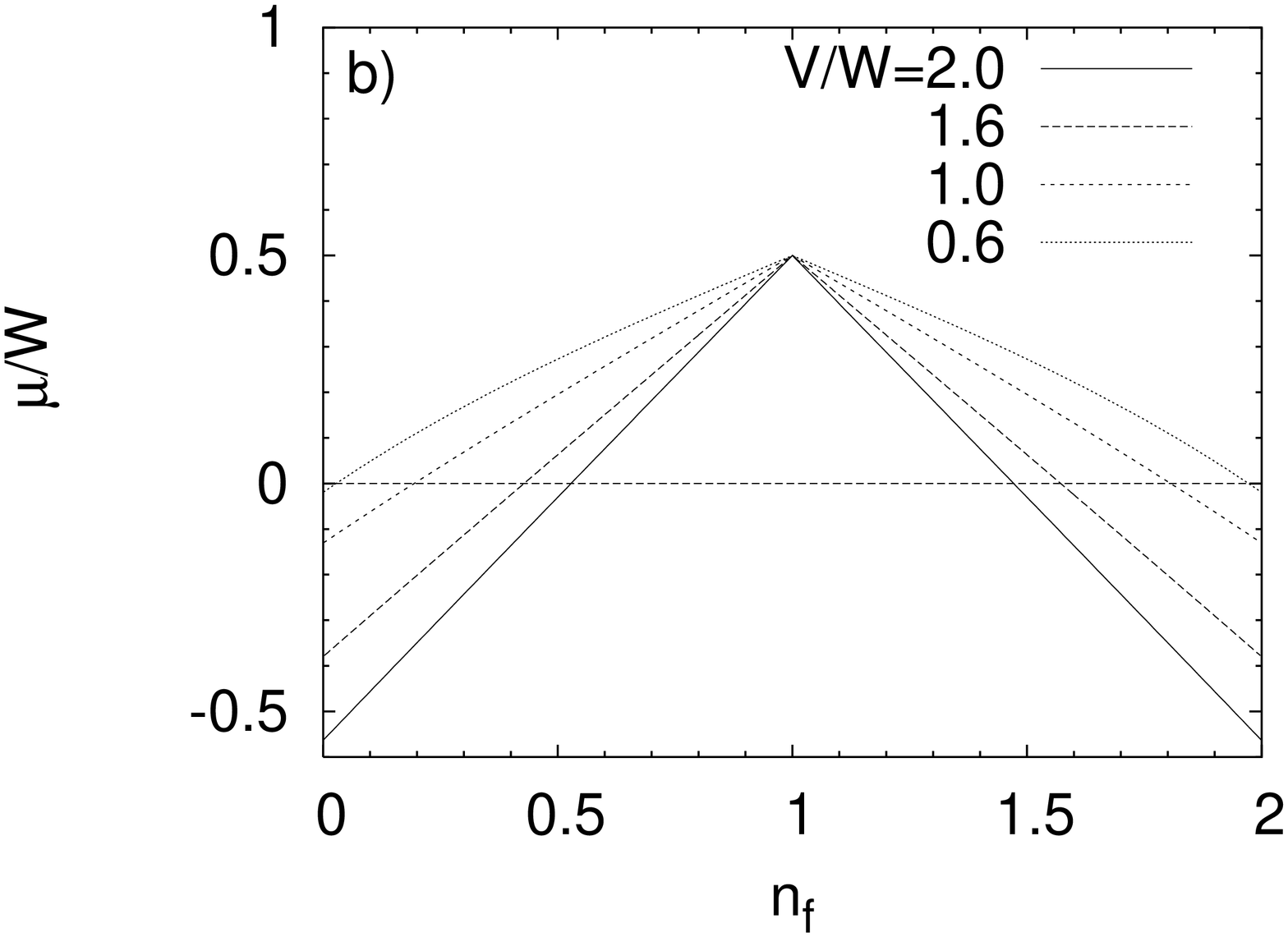}
\caption{The carrier density dependence of 
a) $\Delta_{s}$ and b) $\mu$  at $T=0$
and different values of $V$ in the $s$-wave pairing channel.}
\end{figure}

\begin{figure}[h!]% fig. 2
\centering
\includegraphics[width=2.0in,height=3.0in,angle=270]{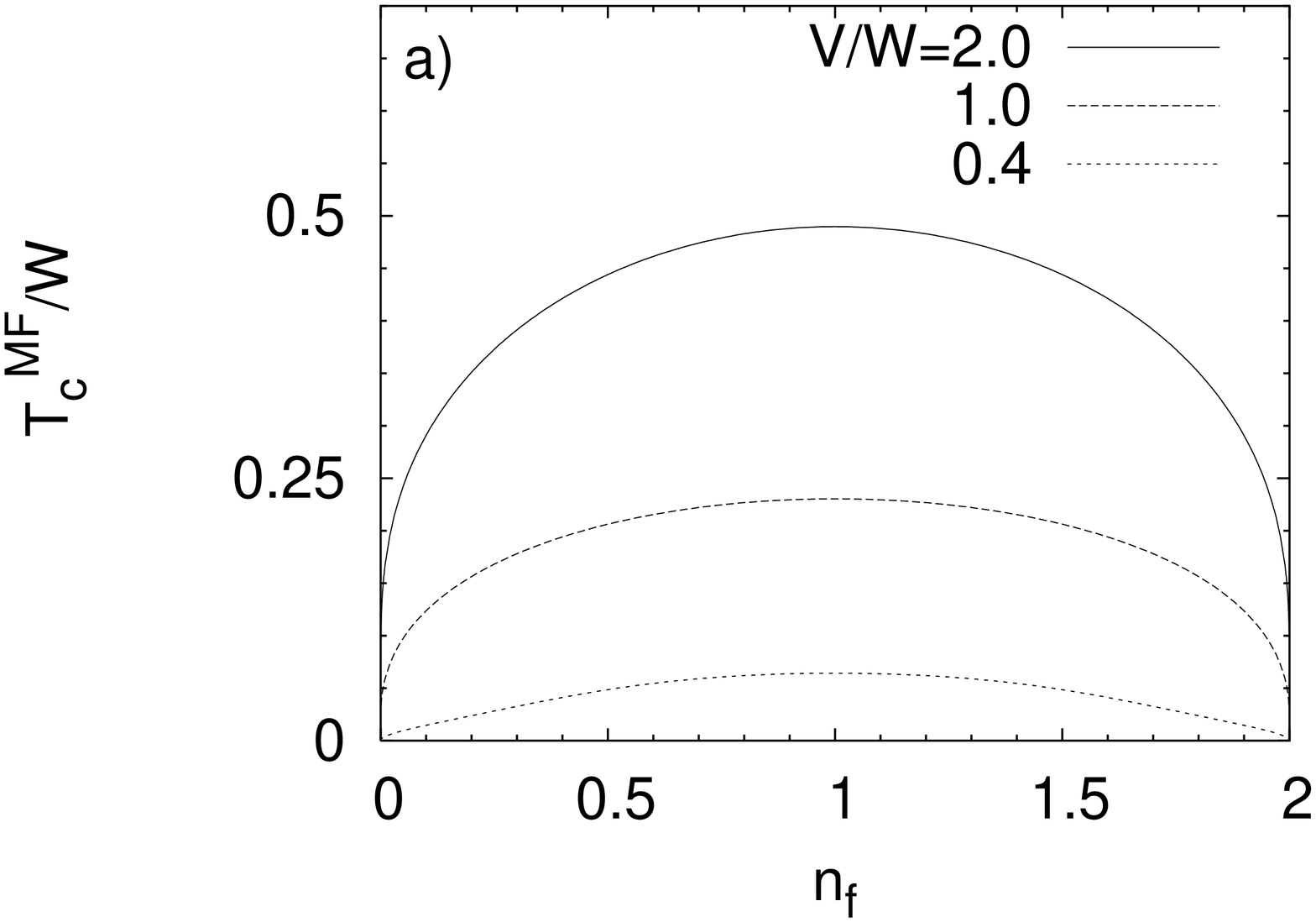}
\includegraphics[width=2.0in,height=3.0in,angle=270]{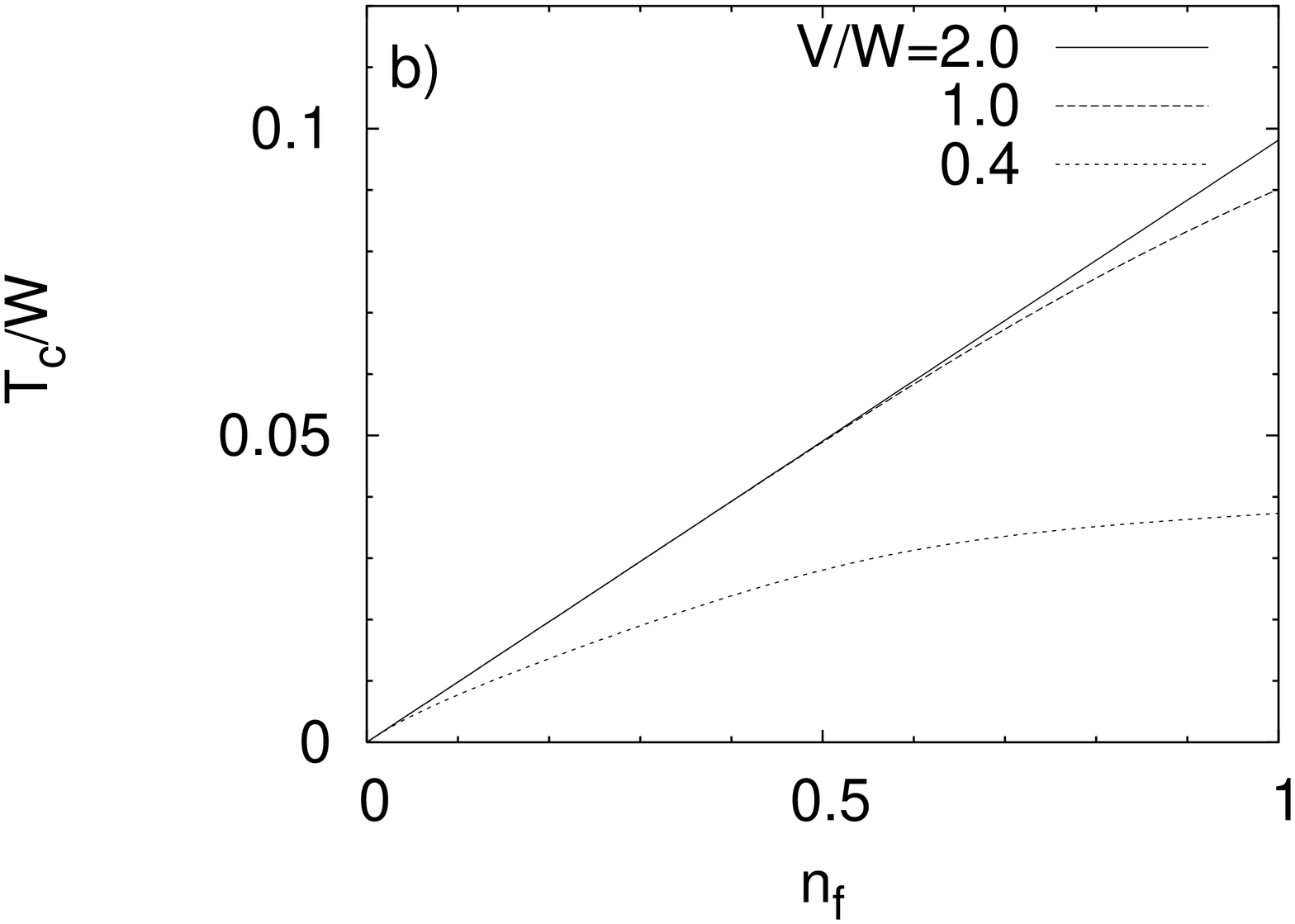}
\caption{The carrier density dependence of 
a) $T_{c}^{MF}$ and b) $T_{c}$ 
at different values of $V$ in the $s$-wave pairing channel.}
\end{figure}

\begin{figure}[h!]% fig. 3
\centering
\includegraphics[width=2.0in,height=3.0in,angle=270]{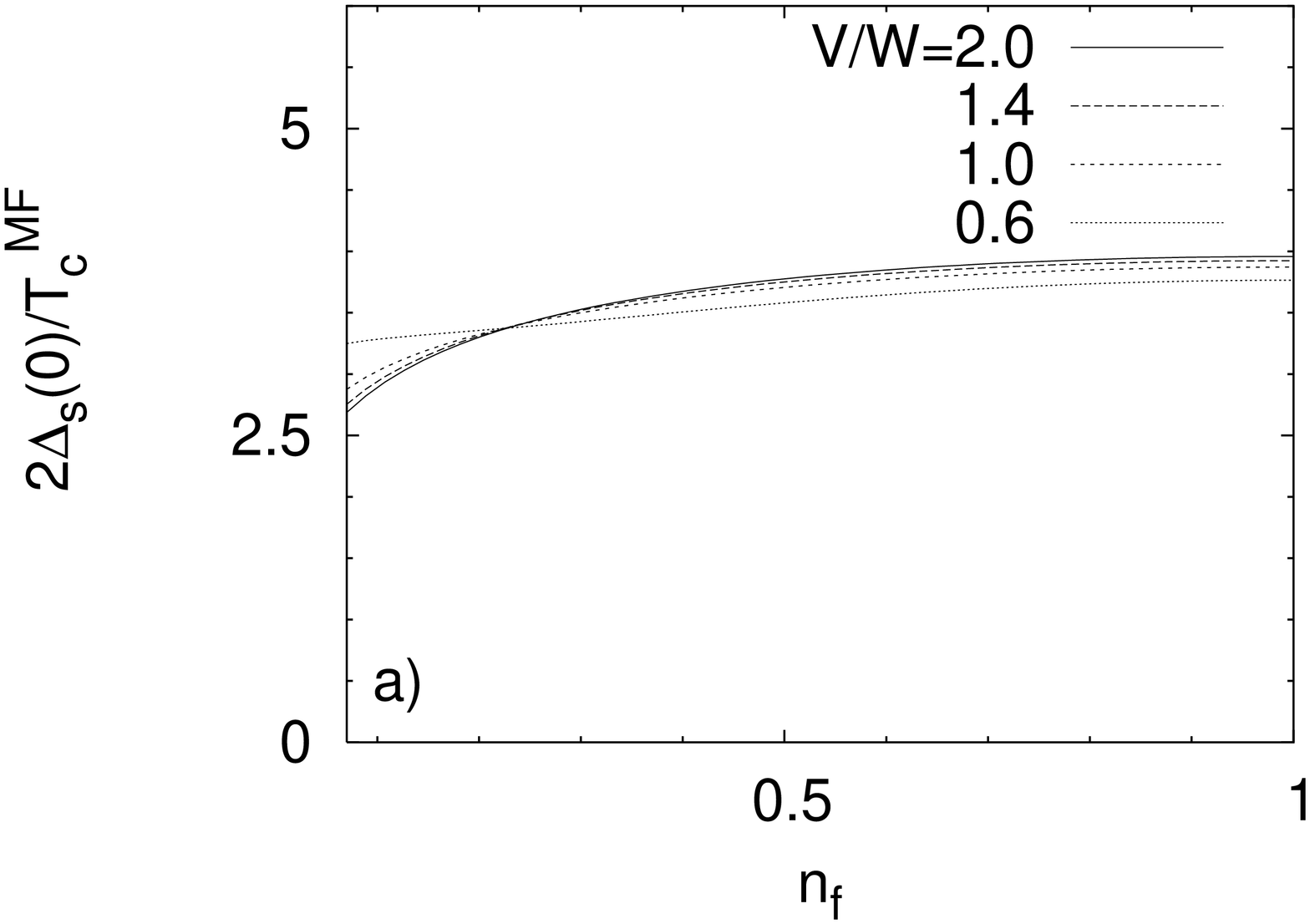}
\includegraphics[width=2.0in,height=3.0in,angle=270]{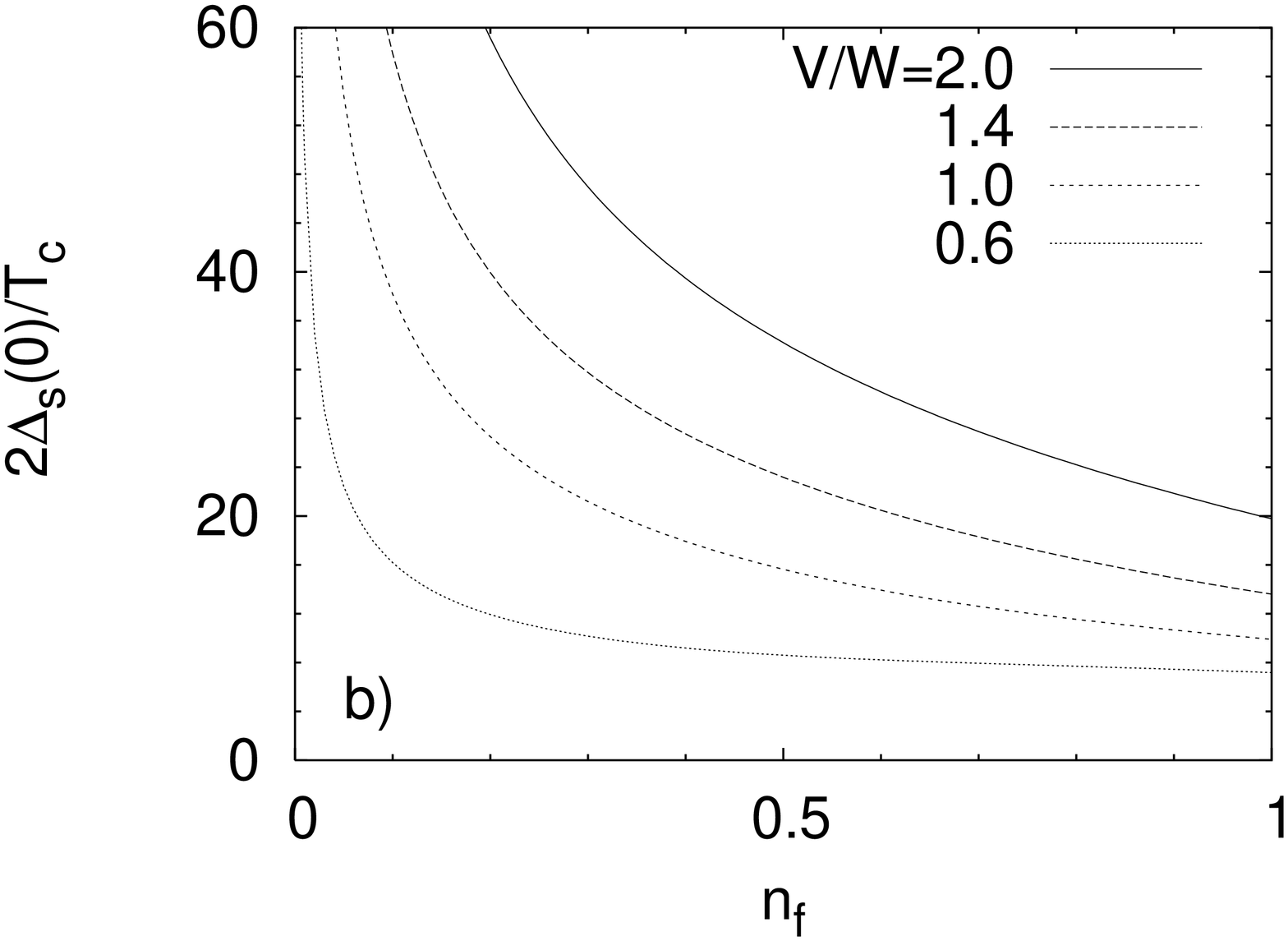}
\caption{The carrier density dependence of the ratios 
a) $2\Delta_{s}(T=0)/T_{c}^{MF}$ (a)) b) $2\Delta_{s}(T=0)/T_{c}$ 
at different values of $V$ in the $s$-wave pairing channel.}
\end{figure}

\newpage

\begin{figure}[h!]% fig. 4
\centering
\includegraphics[width=2.0in,height=3.0in,angle=270]{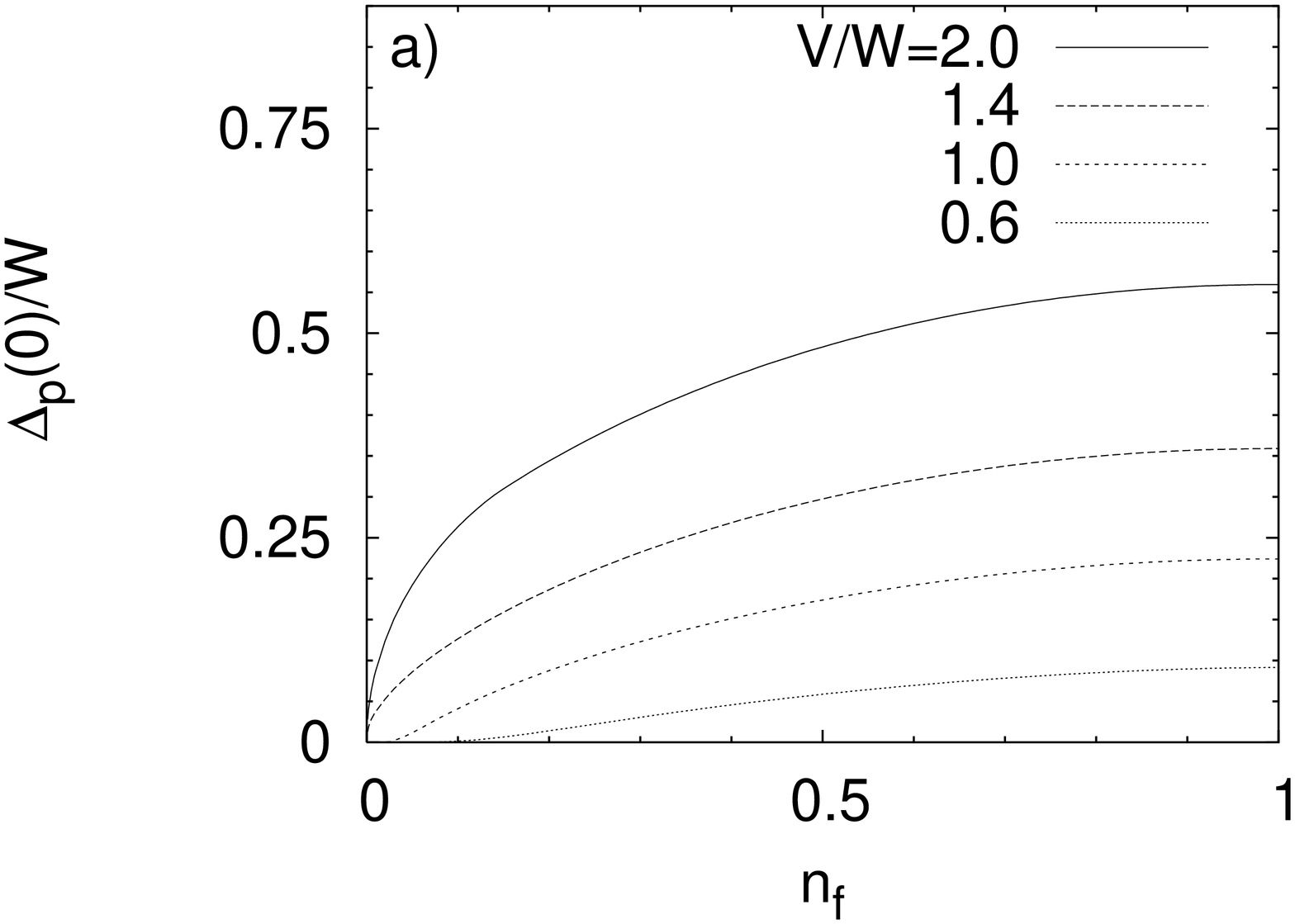}
\includegraphics[width=2.0in,height=3.0in,angle=270]{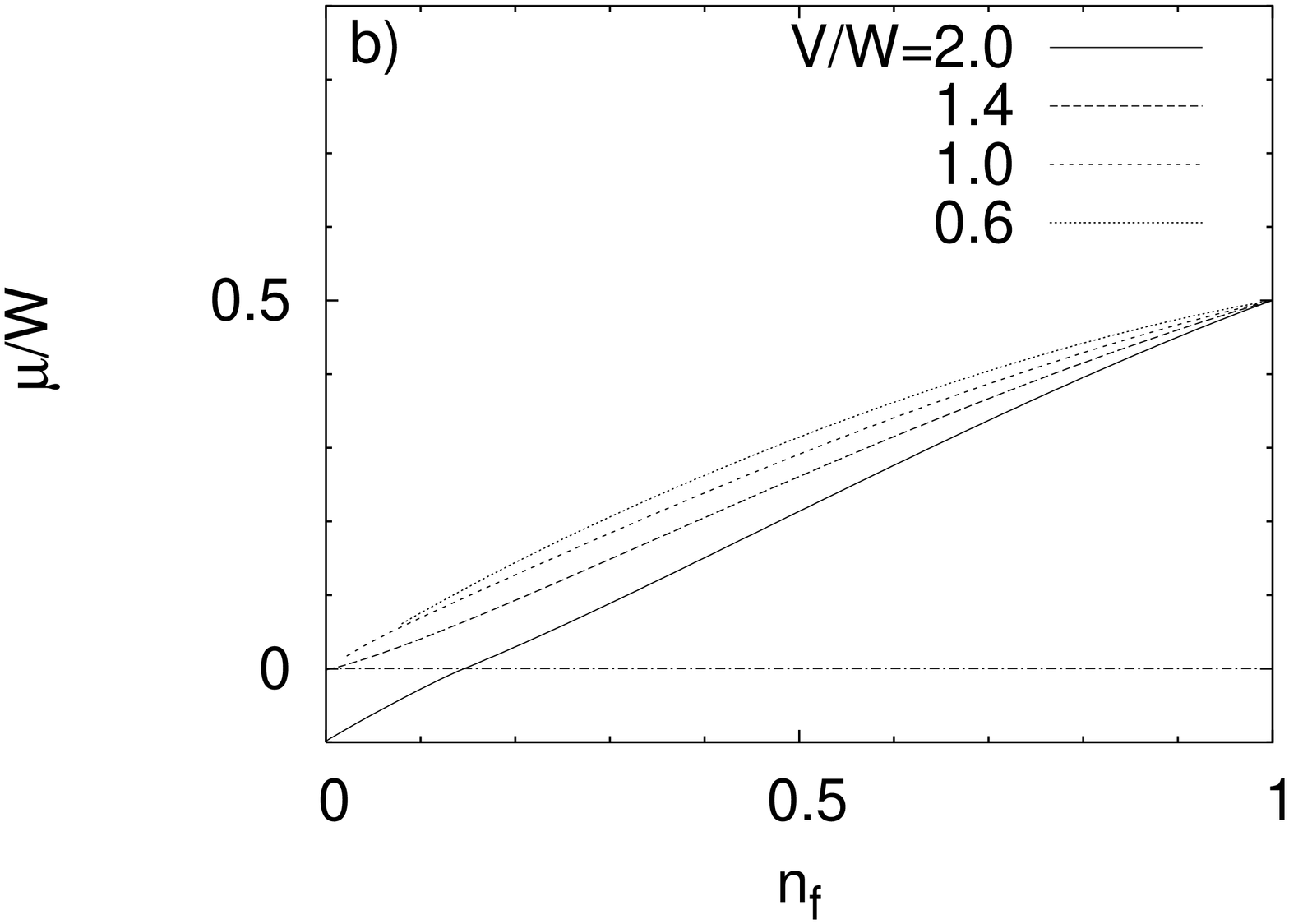}
\caption{The same as in Fig.1 for the $p$-wave pairing case.}
\end{figure}

\begin{figure}[h!]% fig. 5
\centering
\includegraphics[width=2.0in,height=3.0in,angle=270]{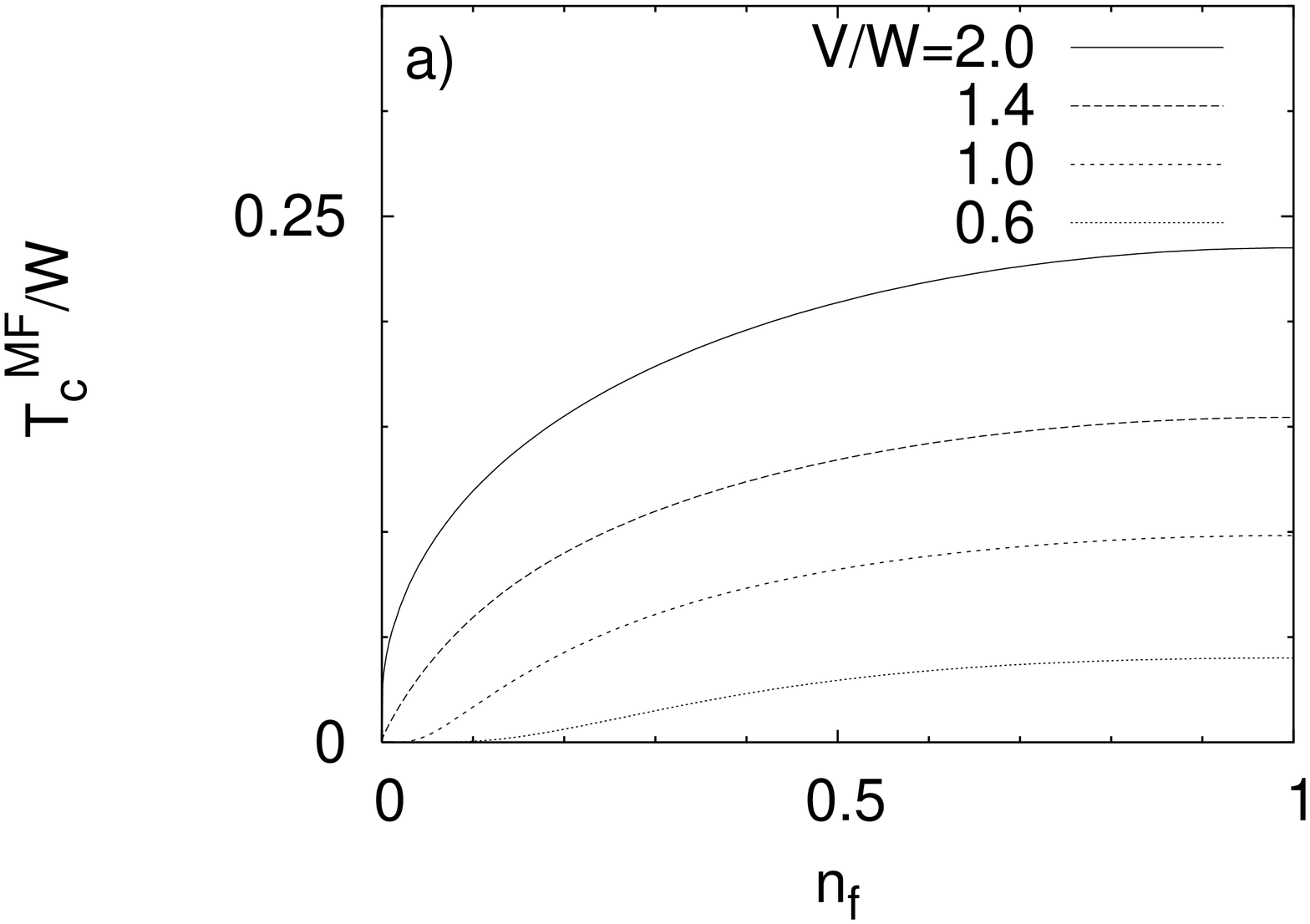}
\includegraphics[width=2.0in,height=3.0in,angle=270]{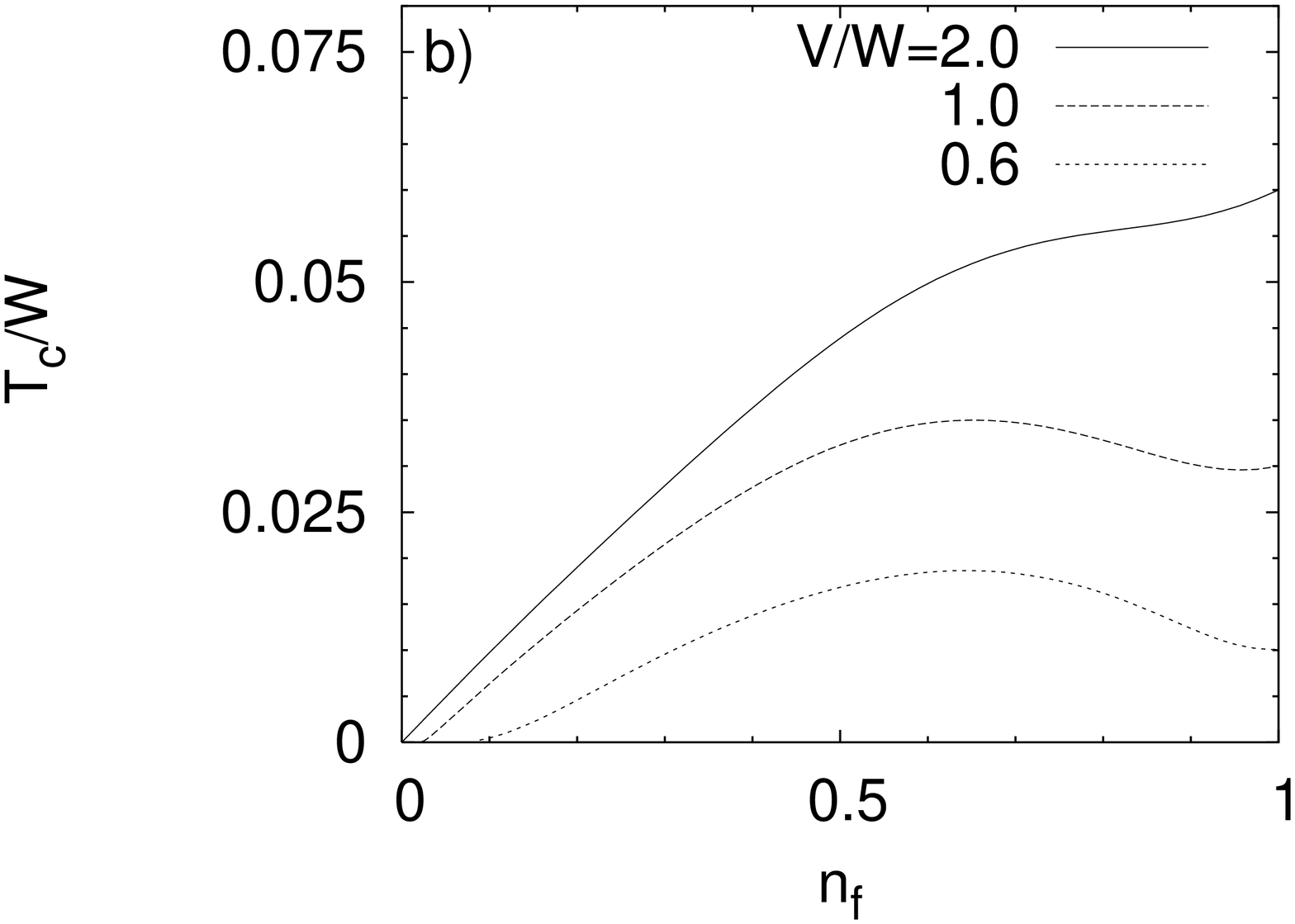}
\caption{The same as in Fig.2 for the $p$-wave pairing case.}
\end{figure}

\begin{figure}[h!]% fig. 6
\centering
\includegraphics[width=2.0in,height=3.0in,angle=270]{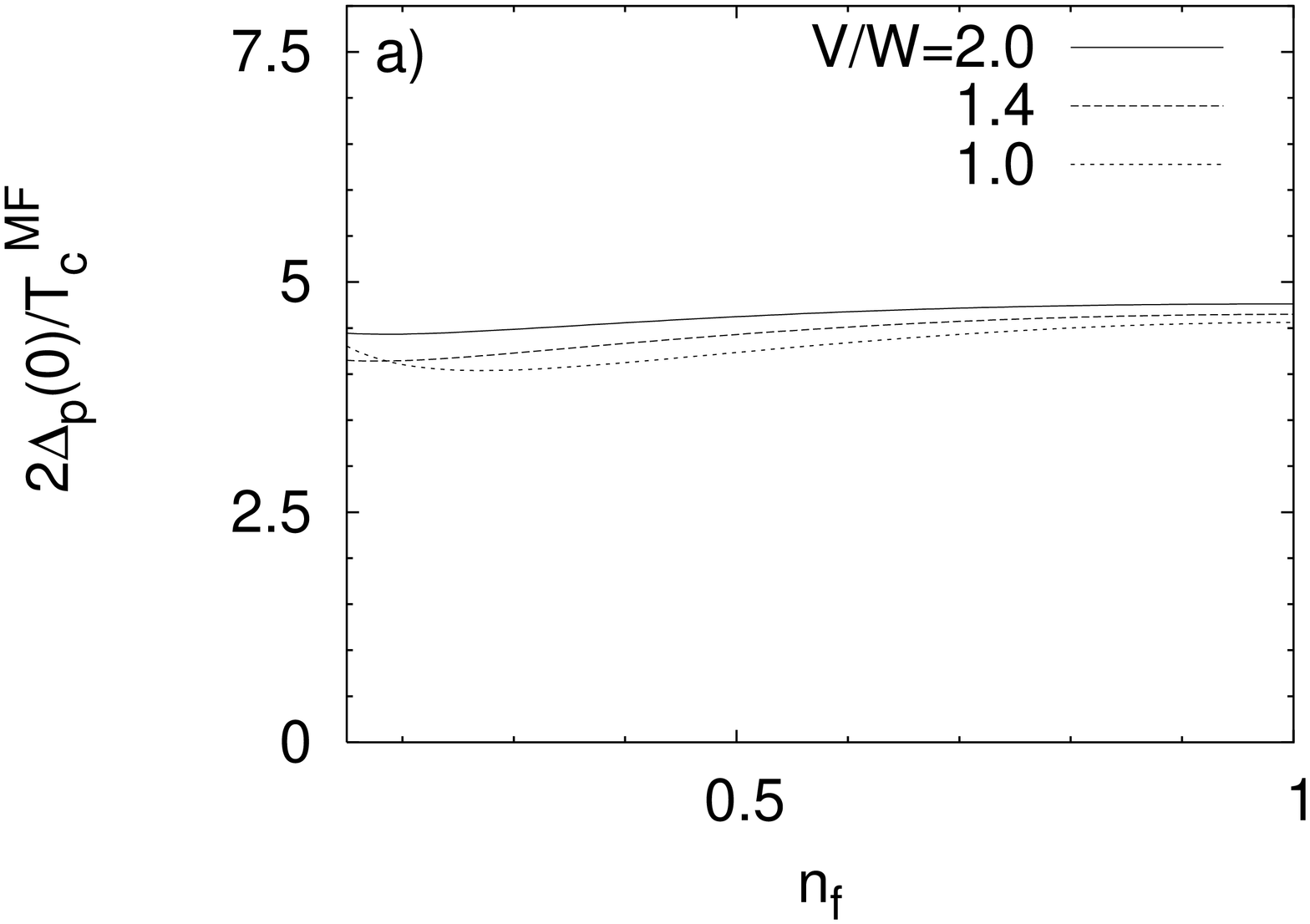}
\includegraphics[width=2.0in,height=3.0in,angle=270]{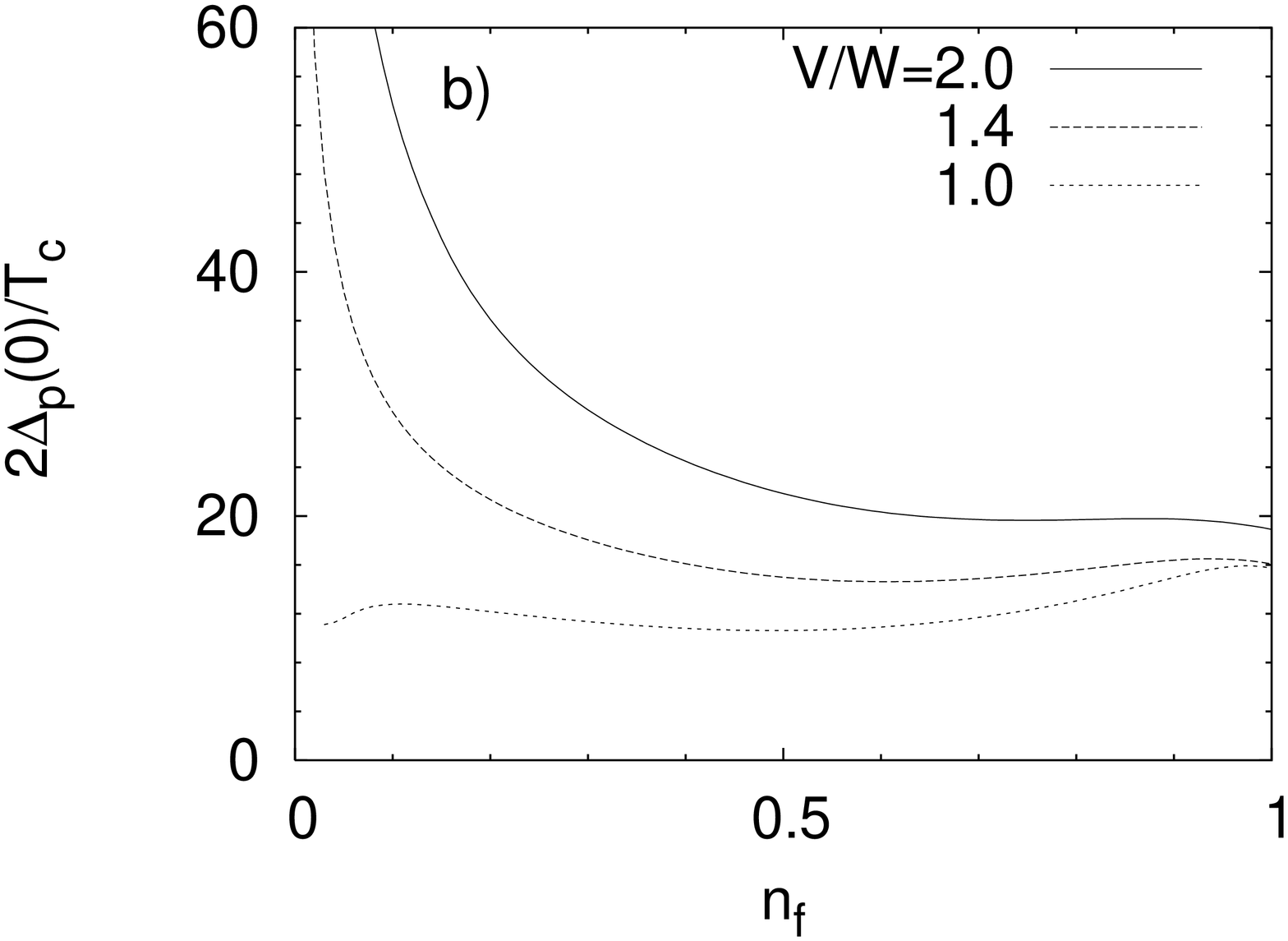}
\caption{The same as in Fig.3 for the $p$-wave pairing case.}
\end{figure}

\newpage

\begin{figure}[h!]% fig. 7
\centering
\includegraphics[width=2.0in,height=3.0in,angle=270]{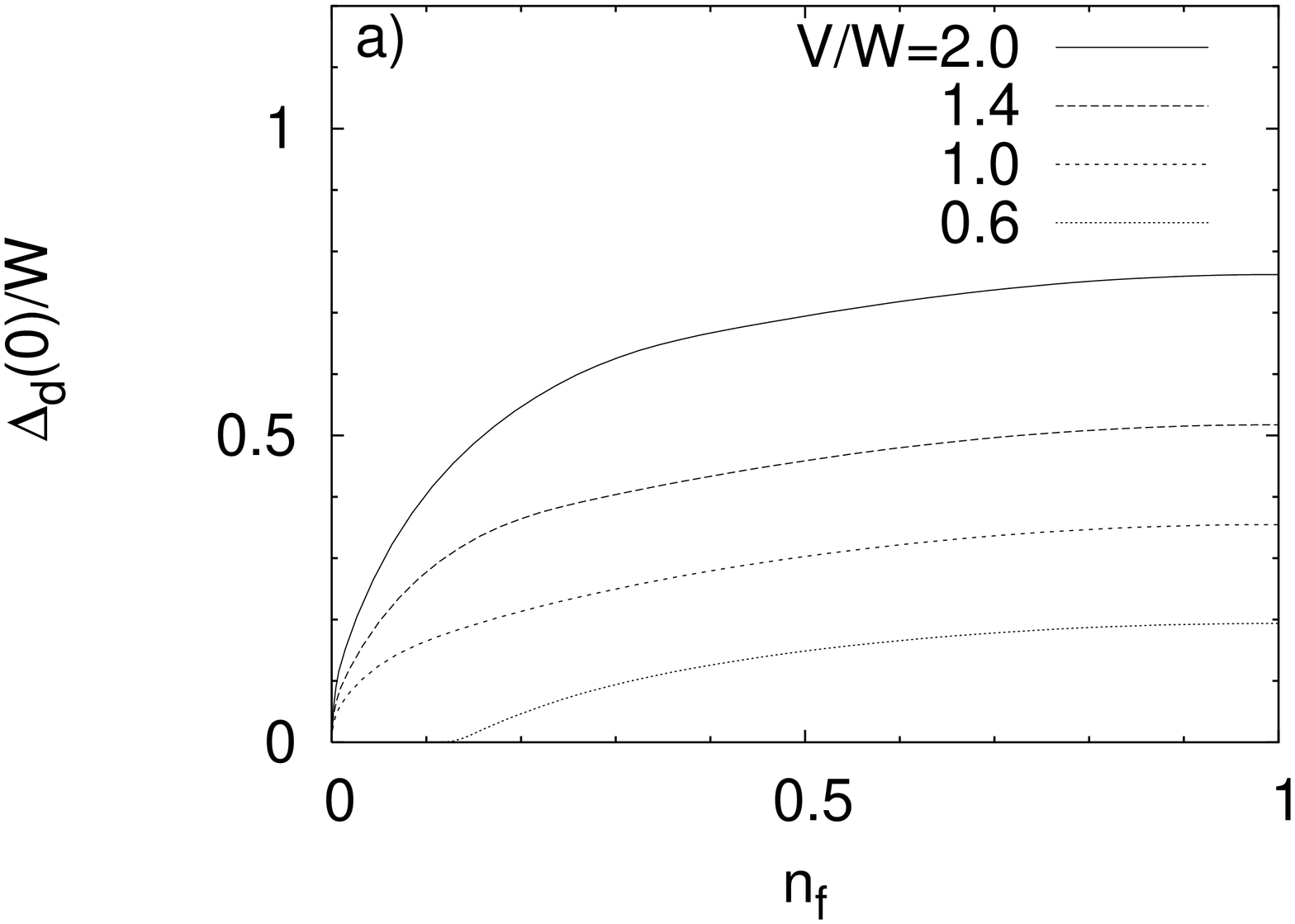}
\includegraphics[width=2.0in,height=3.0in,angle=270]{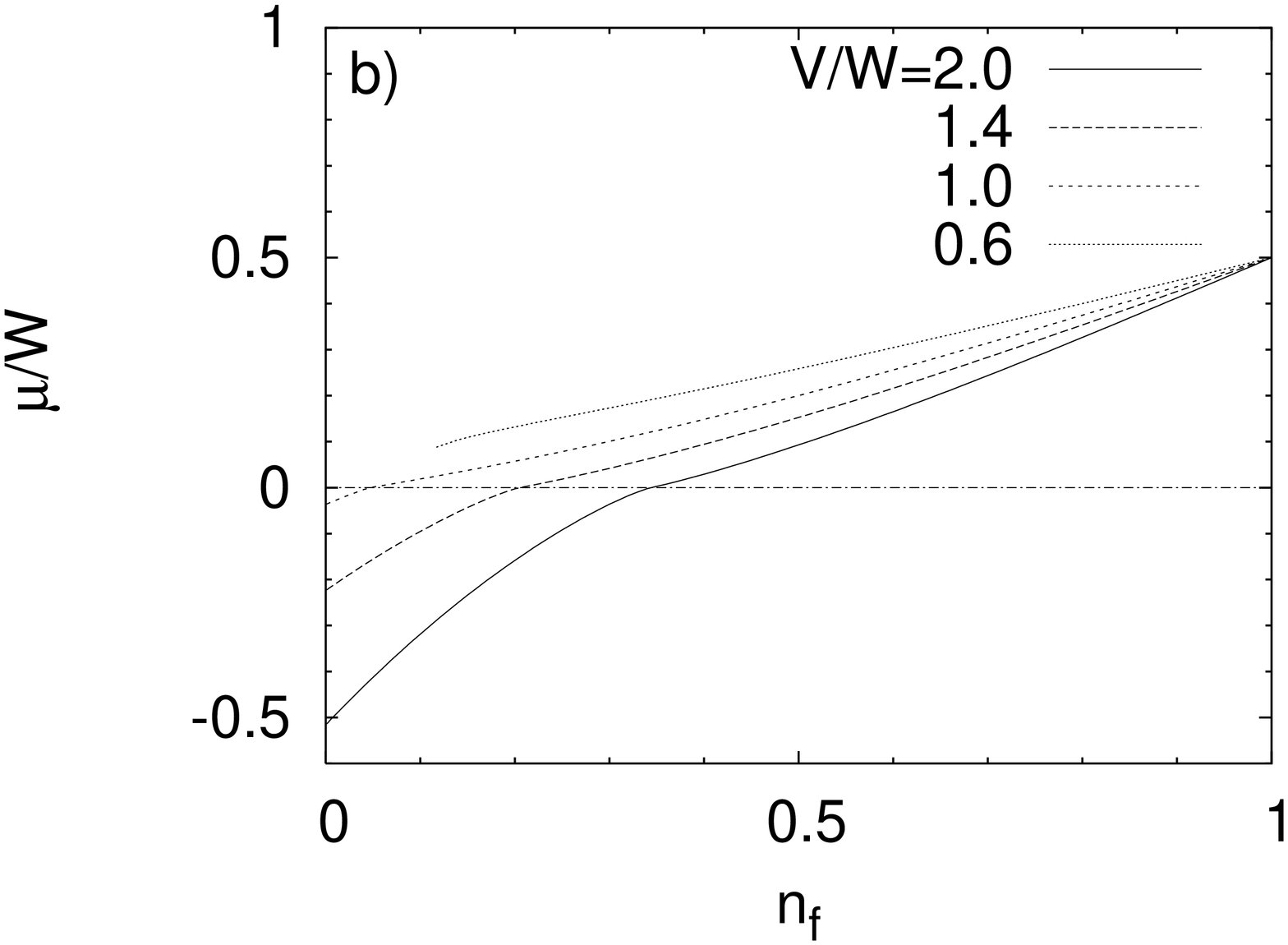}
\caption{The same as in Fig.1 for the $d$-wave pairing case.}
\end{figure}

\begin{figure}[h!]% fig. 8
\centering
\includegraphics[width=2.0in,height=3.0in,angle=270]{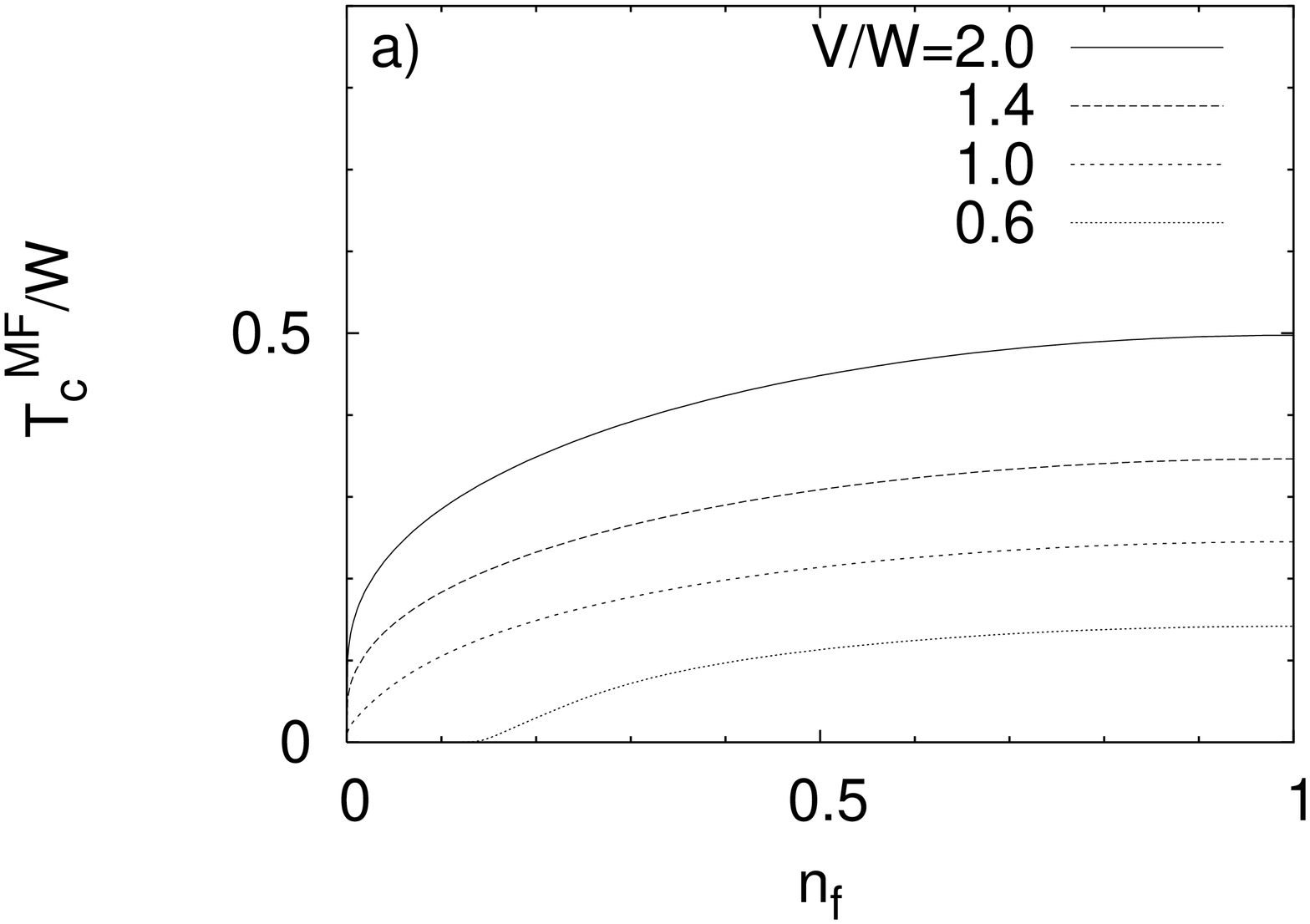}
\includegraphics[width=2.0in,height=3.0in,angle=270]{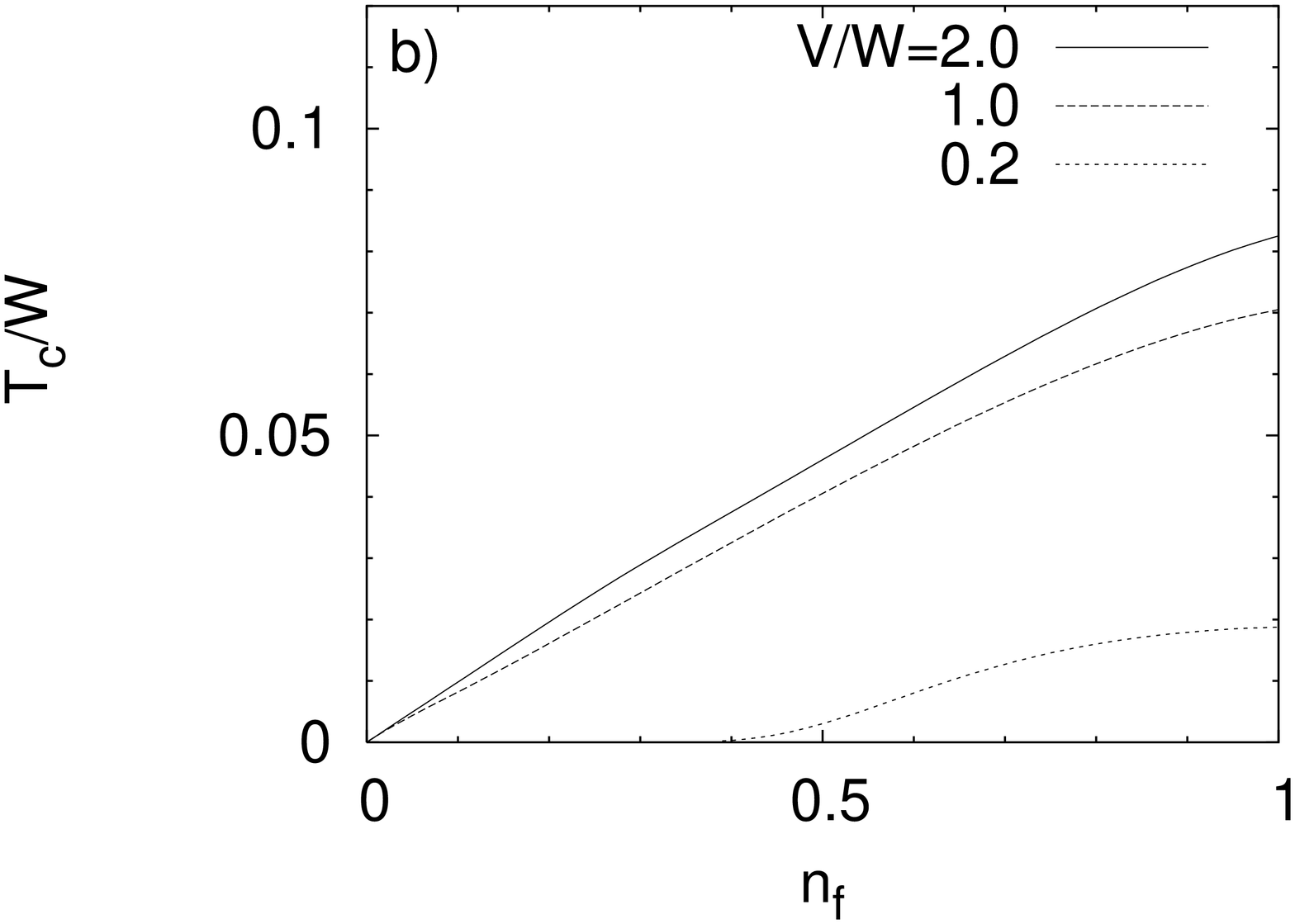}
\caption{The same as in Fig.2 for the $d$-wave pairing case.}
\end{figure}

\begin{figure}[h!]% fig. 9
\centering
\includegraphics[width=2.0in,height=3.0in,angle=270]{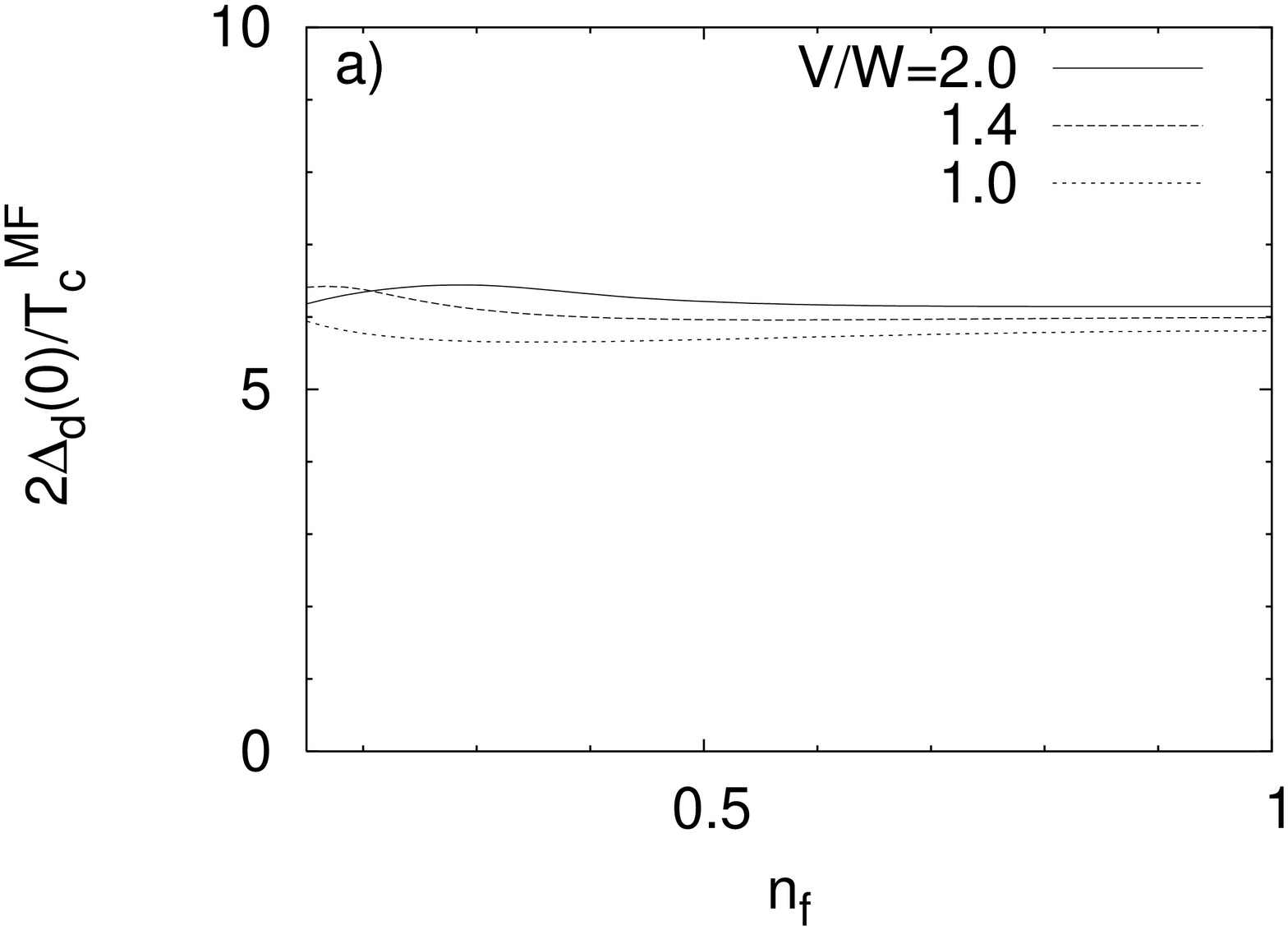}
\includegraphics[width=2.0in,height=3.0in,angle=270]{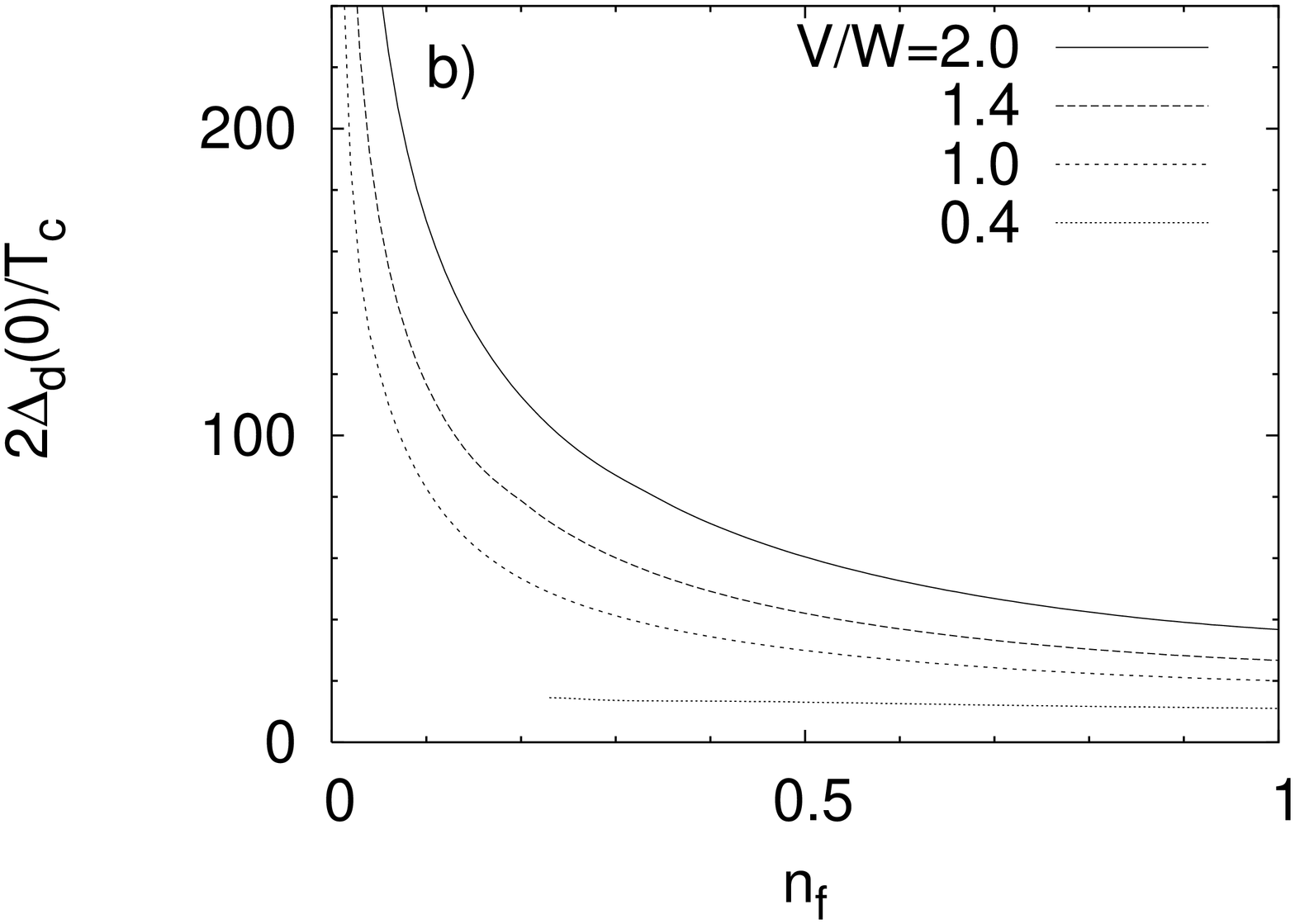}
\caption{The carrier density dependence of the ratios 
a) $4\Delta_{d}(T=0)/T_{c}^{MF}$ and b) $4\Delta_{d}(T=0)/T_{c}$
at different values of $V$ in the $d$-wave pairing channel
(The maximal value of the gap in the $d$-wave pairing
case is $2\Delta_{d}(T=0)$, not $\Delta_{d}(T=0)$ like in the
$s$- and $p$-wave cases.)}
\end{figure}

\begin{thebibliography}{99}
\bibitem{Foot}
It is necessary to mention that usually the temperature
$T_{c}^{MF}$ is not considered in the case of the 2D XY-model.
Being rather high ($T_{c}^{MF}\sim 4J$ for the square
lattice) this temperature shows when and where (above $T_{c}^{MF}$)
the spin system can be considered in the paramagnetic state,
i.e. where the average spin modulus on the lattice is zero.
\end{thebibliography}
\end{document}